# Characteristic ferroelectric domains and their dynamic behavior in ordered Pb(Sc$_{1/2}$Nb$_{1/2}$)O$_3$

Hiroshi Nakajima[1], Satoshi Hiroi[2], Hirofumi Tsukasaki[1], Yonghong Bing[3], Stéphane Grenier[4], Zuo-Guang Ye[3], Pierre-Eymeric Janolin[5], and Shigeo Mori[1]

[1]Department of Materials Science, Osaka Metropolitan University, Sakai, Osaka 599-8531, Japan

[2]Faculty of Materials for Energy, Shimane University, Matsue, Shimane, 690-8504 Japan

[3]Department of Chemistry and 4D LABS, Simon Fraser University, 8888 University Drive, Burnaby, BC, V5A 1S6, Canada

[4]Université Grenoble Alpes, CNRS, Grenoble INP, Institut Néel, 38000 Grenoble, France

[5]Université Paris-Saclay, CentraleSupélec, CNRS, laboratoire SPMS, 91190 Gif-sur-Yvette, France

Pb-based perovskites with multiple cations are fascinating materials showing various phenomena such as high piezoelectric, electromechanical, and relaxor properties. While chemical disordering accompanied by polar nanoregions and nanosized domains is commonly believed to cause the relaxor nature, little is known about ferroelectric microstructures of chemically ordered Pb-based perovskites. In this study, we discovered intriguing meandering ferroelectric domains in chemically ordered ferroelectric Pb(Sc$_{1/2}$Nb$_{1/2}$)O$_3$ using in-situ transmission electron microscopy with dark-field imaging. Observation results demonstrate that electric polarization can fluctuate around the [111] direction despite the formation of long-range ordered rhombohedral domains, which results in unique weak relaxor properties. In-situ imaging upon heating successfully reveals the dynamic behavior of domain-wall movements with lattice distortion and paraelectric–ferroelectric phase coexistence in the vicinity of the Curie temperature, indicating a discontinuous phase transition. Our research provides new insights into the effect of chemical ordering on ferroelectric nanodomains.

**Introduction**

Ferroelectric properties are governed by domain-wall movements and electric-polarization distribution under various external stimuli, such as electric field, mechanical stress, and temperature. In relaxor ferroelectrics, which exhibit frequency dispersion and broad peaks in the temperature dependence of dielectric permittivity [1], decreasing the temperature causes the nonergodic state to freeze into a static ferroelectric nanodomain state. In addition, macroscopic polarization depends on the cooling process in the presence or absence of an electric field, which is expected to result in different microstructures [2, 3]. Direct observation and analysis of ferroelectric domains in real space are essential for understanding the physical and mechanical properties of ferroelectrics. However, ferroelectric domains and their morphology have not been thoroughly characterized at the nanoscale in numerous functional ferroelectrics. In particular, little work has been done to capture the dynamic behavior of ferroelectric domains at phase transitions because of experimental difficulties.

Pb-based perovskites Pb($B'_xB''_{1-x}$)O$_3$ are technologically important because they exhibit large electromechanical responses and high dielectric properties [1]. In particular, Pb(Sc$_{1/2}$Nb$_{1/2}$)O$_3$ (PSN) is a system that is suitable for examining the chemical ordering of $B$ sites that affects its ferroelectric properties [4, 5]. If the differences in the radius and charge of $B$-site cations are sufficiently large, a superstructure of alternate cation elements is realized along [111]$_{pc}$ (pc: pseudocubic), which results in normal ferroelectricity. Meanwhile, the chemical disordering of $B$ sites causes relaxor ferroelectricity with characteristic dielectric dispersion. The difference in cation size in PSN is close to the boundary between chemical ordering and disordering: The ionic sizes of Sc$^{3+}$ (0.75 Å) and Nb$^{5+}$ (0.64 Å) provide a radius difference of 15.8%, which falls into the critical point range between 7% and 17% for an ordered superstructure to occur in Nb-based perovskites [5, 6]. This enables the control of the relaxor/nonrelaxor properties through thermal treatments, such as quenching and annealing [7–9]. Disordered PSN exhibits relaxor ferroelectricity, whose temperature of maximum permittivity depends on the degree of ordering. Similar to other relaxor ferroelectrics, the relaxor properties of disordered PSN are related to nanosized ferroelectric domains, i.e., polar nanoregions [10–13]. In fact, the polar nanoregions of disordered PSN are visualized as nanodomains, and these domains dynamically change into lamellar-like and irregularly shaped ones at the ferroelectric–relaxor phase transition [11, 14].

Conversely, $B$-site ordered PSN exhibits normal ferroelectricity with a rhombohedral structure [6, 15, 16]. However, some peculiar behaviors were reported in its dielectric properties; a weak dielectric relaxation developed even in a highly ordered specimen with a degree of ordering of $S = 0.92$ [15]. Note that the degree of ordering $S$ is defined by the ratio of the integrated intensities of the 1/2 1/2 1/2 and 1 0 0 peaks divided by the same ratio calculated for a perfectly ordered structure: $S = \left(\frac{I_{1/2\ 1/2\ 1/2}}{I_{1\ 0\ 0}}\right)^{\frac{1}{2}}_{\exp} / \left(\frac{I_{1/2\ 1/2\ 1/2}}{I_{1\ 0\ 0}}\right)^{\frac{1}{2}}_{\text{cal}}$. Furthermore, a clear deviation from the Curie–Weiss law was observed at 500 K, which is far above the Curie temperature of ordered PSN ($T_C \approx 345$ K) [16]. These results show some unique relaxor-like characteristics, i.e., weak relaxor properties, of ferroelectric PSN. Although the intriguing ferroelectric nature was reported, as explained earlier, the behavior and morphology of ferroelectric domains in ordered PSN have not been revealed, unlike in numerous reports of disordered PSN.

In this study, we unveil intriguing structures of ferroelectric domains in ordered PSN using dark-field imaging and in-situ heating by transmission electron microscopy (TEM). Observation results point to the characteristic ferroelectric morphology inherited from the rhombohedral structure in PSN. The ferroelectric domains have small meandering shapes of a few hundred nanometers. We further demonstrate the real-space dynamics of the domain walls at the ferroelectric phase transition. The in-situ observation reveals paraelectric–ferroelectric phase coexistence and captures domain-wall movements in the vicinity of the Curie temperature.

**Methods**

TEM was performed using a JEM-2100F accelerated at 200 kV (JEOL Co. Ltd.). Dark-field imaging was used to visualize the ferroelectric domains by tilting the specimen under two-beam conditions. This technique depicts polar-dependent domains due to the breakdown of Friedel's law [17, 18]. The specimen temperature was controlled using a heating double-tilt holder (JEOL Co. Ltd.). The dynamics of the domain change were recorded at a rate of 0.05 s/frame on a CMOS camera with an in-situ mode (Gatan OneView). The vacuum pressure of the JEM-2100F instrument was $1.0 \times 10^{-5}$ Pa. Single crystals of $B$-site ordered PSN were grown using a high-temperature solution method [19–21]. Polycrystalline specimens of $0.80Pb(Sc_{1/2}Nb_{1/2})O_3$–$0.20PbTiO_3$ and $0.91Pb(Sc_{1/2}Nb_{1/2})O_3$–$0.09PbTiO_3$ were synthesized using a solid-state reaction method [22]. The single crystal was cut parallel to the (001) plane and thinned via mechanical polishing, followed by Ar ion milling. The surfaces of the specimens were coated with carbon to prevent electron charging. Without the carbon deposition, the ferroelectric domain structures were found to be changed from the pristine state and transformed into microsized zigzag shapes, as shown in Supplementary Fig. S1.

The temperature dependence of powder XRD was measured at the European Synchrotron Radiation Facility using the D2AM beamline, whose incident X-ray beam energy was 15.197 keV. Local structures were determined using pair-distribution function (PDF) analysis at the BL04B2 beamline (SPring-8) [23, 24], whose incident X-ray beam energy was set to be 113 keV to reduce X-ray absorption due to the presence of Pb. The experimental PDF $G(r)$ was calculated as follows:

$$G(r) = \frac{2}{\pi} \int_{Q_{\min}}^{Q_{\max}} Q[S(Q) - 1] \sin(Qr) \, dQ,$$

where $Q_{\max} = 24.7$ Å$^{-1}$ ($Q_{\min} = 0.3$ Å$^{-1}$) is the maximum (minimum) value of the scattering vector and $S(Q)$ is the structure factor, which is obtained by normalizing the experimental coherent scattering intensity [25]. The ground specimen was packed into a quartz capillary with an inner diameter of 0.3 mm, and the temperature was controlled using a vacuum electric furnace.

**Results and discussion**

To obtain an insight into the crystal structure of PSN, we investigated the temperature dependence of the XRD patterns. Figure 1(a) shows the temperature dependence of the Bragg reflection $4\bar{4}4$. The peak splits at approximately 340 K upon cooling, indicating the onset of rhombohedral distortion from the simple cubic structure. Previous studies have demonstrated that the transition temperature decreases with increasing degree of ordering from 400 K ($S = 0$) to 350 K ($S = 0.92$) [8, 15]. Hence, the temperature of 340 K indicates that the degree of ordering $S$ of the single crystal studied in this work is higher than 0.90.

The PDF analysis results further reveal the rhombohedral distortion, as shown in Fig. 1(b). Because the unit cell of ordered PSN is $2a_{pc} \times 2a_{pc} \times 2a_{pc}$ compared with that of the disordered structure ($a_{pc}$: lattice constant) [26], the displacement patterns of the individual ions are highly complex, which prevented us from refining the structure. However, Fig. 1(b) shows that with decreasing temperature, the second (5.8 Å) and fourth (9.08 Å) peaks are broadened, while the first peak at 3.0–4.3 Å does not change. The first peak corresponds to the pseudocubic unit cell, while the second and fourth peaks are related to the diagonal spacing. This indicates that the nearest-neighbor Pb–Pb ion distances are likely to remain unchanged and that the second-nearest-neighbor Pb–Pb ion distances tend to vary, as shown in Supplementary Fig. S2(a). These phenomena are consistent with the effect of diagonal distortion.

Another noticeable feature is that upon cooling from 600 K to 300 K, the intensity of the fifth peak (9.98 Å) increases while that of the sixth peak (10.40 Å) decreases. Simulation of the PDF analysis was performed to understand the structural change associated with this trend [Supplementary Fig. S2(b) and (c)]. When a uniform rhombohedral distortion is introduced to deviate the angle $\alpha$ from 90° in the perovskite structure, the PDF peaks show the opposite trend with the intensity of the sixth peak increasing. This may indicate that the Nb(Sc) ions of the Pb–Nb pairs (i.e., sixth peak), which are described in Supplementary Fig. S2(d), could be shifted from the (1/2, 1/2, 1/2) position to increase their atomic correlation in the rhombohedral structure. Although the XRD results can clarify the lattice distortion in PSN, characterizing the local ferroelectric domains requires microscopy techniques.

Figure 2 shows the electron diffraction pattern and dark-field images of PSN. The diffraction pattern in Fig. 2(a) indicates the (001) plane of the observation area. Within the resolution, the rhombohedral distortion from the cubic structure is too small to detect. The dark-field images show the ferroelectric domains originating from the breaking of inversion symmetry. Under two-beam conditions, the ferroelectric domains generally appear bright (dark) when the electric polarization is parallel (antiparallel) to the Bragg reflection used in dark-field imaging [27, 28]. The contrast disappears when the electric polarization is perpendicular to the Bragg reflection. The dark-field images using $g_{\bar{1}\bar{1}0}$ and $g_{110}$ show a characteristic wavy pattern. The contrast is reversed in the images using $g_{\bar{1}\bar{1}0}$ and $g_{110}$, as shown in Fig. 2(b) and (c). Conversely, it disappears in the images using $g_{1\bar{1}0}$ and $g_{\bar{1}10}$ (Supplementary Fig. S3). These results demonstrate that the contrast indeed originates from the ferroelectric domains. The projected polarization points along the $[\bar{1}\bar{1}0]_{pc}$ or $[110]_{pc}$ direction, as indicated by the blue and red arrows in the observed plane.

Different directional domains exist in the neighboring areas. The dark-field image in Fig. 2(d) reveals the ferroelectric domains with the polarization component along $[1\bar{1}0]_{pc}$, unlike in Fig. 2(b) and (c). The contrast disappears in the area with the polarization component along $[110]_{pc}$, and vertical directional domains are formed across the boundary. Because the TEM images are projected along the [001] direction, the upward or downward direction of polarization cannot be distinguished, and only the [110] component of the [111] polarization can be identified for such a zone axis. Nonetheless, the formation of four ferroelectric domains agrees with the rhombohedral $R3m$ structure characterized by the polarization oriented along the four $<111>_{pc}$ directions, as shown in Fig. 2(e). Notably, the observation results indicate that the ferroelectric domains of two polarization directions are paired, while those of the other polarization directions are formed in another twin domain.

A noticeable characteristic of the observed domains is the wavy shapes of the ferroelectric domains. The striped domains observed in Fig. 2 are similar to the ferroelectric domains found in BiFeO$_3$, which has the rhombohedral structure $R3c$, except that those in BiFeO$_3$ are straight in shape [29]. Conversely, the ferroelectric domains in PSN exhibit meandering structures and Y-shaped curves. These results demonstrate that the electric polarization is rotated at the domain walls and that the local polarization direction can fluctuate around the $<111>_{pc}$ axis, which may be due to the weak anisotropy of the electric polarization in PSN. The XRD analysis revealed that the coherence length of the chemical ordering ranges from 20 to 80 nm even in highly ordered PSN and Pb(Sc$_{1/2}$Ta$_{1/2}$)O$_3$ with $S \geq 0.92$ [7, 15]. Thus, the observed specimen could show local inhomogeneity over several tens of nanometers, which would cause the weak anisotropy of the polarization direction. The weak anisotropy should make the domain size smaller in PSN. Notably, similar wavy patterns were observed in BaTiO$_3$ and triglycine sulfate, which exhibit particular macroscopic polarization directions without cation disordering [30–32]. However, the domain sizes of these materials are micrometers, while that of PSN is a few hundred nanometers. Such a small meandering ferroelectric domain structure could be related to the weak relaxor properties of PSN because the fluctuation of the local static polarization similar to relaxors could affect its dielectric responses exhibiting a relaxor-like behavior, even though the long-range ordered ferroelectric property is dominant [15, 16]. Furthermore, this TEM study using ordered ferroelectric PSN with $S > 0.90$ indicates that decreasing the degree of ordering makes the size of ferroelectric domains smaller and eventually results in polar nanoregions, which were visualized using piezoresponse force microscopy [14]. Simultaneously, it causes the measured relaxor properties [1].

To investigate the thermal evolution of the ferroelectric domains, in-situ observation was performed over a wide temperature range. Figure 3 shows the domain structures observed from the same area upon changing the temperature. When the specimen was heated, the domain structure remained unchanged below 330 K, as shown in Fig. 3(a) and (b), indicating that the domain structure is robust against temperature change when the temperature is 12 K below the Curie temperature. Increasing the temperature reduces the domain contrast because the magnitude of the electric polarization becomes smaller without structural changes. This behavior is consistent with normal ferroelectricity and does not agree with relaxor ferroelectrics because the size of polar nanodomains typically decreases with increasing temperature in relaxors [24, 33]. At 343 K, the ferroelectric domains disappeared, as shown in Fig. 3(c), revealing a ferroelectric phase transition at the Curie temperature $T_C = 342$ K. The Curie temperature agrees with the onset of the rhombohedral distortion observed in the XRD pattern in Fig. 1 and the previous result on highly ordered single crystals [6]. Notably, although the deviation from the Curie–Weiss law appeared at a temperature above 500 K [16], the TEM observation shows no nanosized ferroelectric domains above $T_C = 342$ K. Thus, the deviation is likely to result from atomistic interactions due to the local inhomogeneity in the crystals but not from the polar nanoregions formed in relaxor ferroelectrics.

When the specimen is cooled down below $T_C$, the ferroelectric domains appear again at 338 K [Fig. 3(d)]. Further cooling the specimen down to 293 K increases the contrast of the domains, but the domain structure remains mostly unchanged compared with that at 338 K, as shown in Fig. 3(e). Thus, the domains do not move between room temperature and 338 K, a temperature several degrees lower than the Curie temperature. The contrast in the central area disappears at 293 K in Fig. 3(e) with $g = \bar{1}\bar{1}0$, while domain patterns that are similar to the pristine state are formed in the right region across the yellow line. The dark-field image with $g = 1\bar{1}0$ reveals the detailed contrast in the central area, as shown in Fig. 3(f). Thus, the central area has the electric-polarization component along $[1\bar{1}0]_{pc}$, demonstrating that thermal annealing causes an electric-polarization flip and changes the domain structure to a structure different from the original one. Another set of ferroelectric domains is formed in the central areas, and the yellow line in Fig. 3(e) shows the boundary between the two sets of domains of different polarization directions.

Close monitoring of the critical phenomenon in the vicinity of the Curie temperature reveals the nature of the ferroelectric phase transition. Figure 4 shows a series of dark-field images with changing temperature (see the Supplementary Movie). When the temperature increases from 338 K to 342 K, the boundary between the ferroelectric and paraelectric regions is moved, as shown by the yellow line, while the magnitude of the electric polarization remains nearly unchanged. Further increase in the temperature enlarges the paraelectric region, and finally, the ferroelectric regions disappear at $T_C$. When the boundary is shifted, the black lines of the bend contours move along with the ferroelectric–paraelectric phase boundary. Because the bend contours are evidence of strain, this indicates that the phase transition is accompanied by the removal of lattice distortion from the crystal.

When the specimen is cooled down from the paraelectric phase, the ferroelectric domains reappear, as shown in Fig. 4, at temperatures from 341 K to 338 K. Notably, the area marked with the yellow arrowhead shows no contrast at 340 K, indicating that the area is a paraelectric state surrounded by ferroelectric domains. Meanwhile, the area gains ferroelectric contrast at 339 K, which shows a ferroelectric phase transition at the nanoscale. Note that the absence of contrast indicates a paraelectric phase or a ferroelectric domain with the other directions of electric polarization. However, the area without contrast would be paraelectric because flipping the polarization directions requires large energy, such as the application of an electric field. A similar nanoscale transition is observed in the area marked with the arrowhead at 338 K. This phenomenon indicates that the ferroelectric domains coexist with the paraelectric regions, which agrees with a discontinuous phase transition with a small temperature interval rather than a diffuse phase transition of a relaxor.

Finally, comparing the results observed in Fig. 2 with those of other materials provides deeper insights into the morphology of the ferroelectric domains in PSN. When PSN is substituted by $PbTiO_3$ to form the $Pb(Sc_{1/2}Nb_{1/2})O_3$–$x$$PbTiO_3$ (PSN–$x$PT) solid solution, a rhombohedral $R3m$ structure exists when $0 < x < 0.3$ [34]. Figure 5 shows the ferroelectric domains observed in 0.80PSN–0.20PT using dark-field imaging, which demonstrates microscale ferroelectric domains with straight domain walls. The presence of electric polarization can be verified in these domains by showing the contrast reversal and disappearance when different Bragg reflections are used, as shown in Fig. 5(a)–(d). The contrast variation demonstrates four different domains, as illustrated in Fig. 5(e). The electron diffraction pattern in Fig. 5(f) indicates that these domains have both the $<110>_{pc}$ and $<001>_{pc}$ components of electric polarization, which agrees with the rhombohedral structure. In addition, similar macroscopic straight ferroelectric domains are observed in 0.91PSN–0.09PT (see Supplementary Figs. S4 and S5). The substitution of $PbTiO_3$ should increase the anisotropy that the electric polarization tends to align along a particular crystallographic axis in PSN–$x$PT. In fact, simulation studies using the Ginzburg–Landau equation indicate that meandering domains are transformed into straight domains when the difference in the gradient coefficients between the [100]- and [010]-polarization components becomes large in the two-dimensional free energy [35, 36]. In addition, the domain size becomes large when the magnitude of the gradient coefficients increases. These results demonstrate that the pristine chemically ordered PSN has a weak anisotropy of polarization and that the electric fields caused by electron radiation or Ti substitution can change the stable polarization configurations in terms of free energy, as shown in Fig. 5 and Supplementary Fig. S1. Interestingly, although both PSN and PSN–$x$PT have similar rhombohedral $R3m$ structures, their ferroelectric domains exhibit different morphologies and sizes. The small meandering rhombohedral ferroelectric domains represent the unique nature of PSN that shows both ferroelectricity and weak relaxor behavior.

## Summary

We investigated the nature of ferroelectric domains and the dynamics of phase transition in $B$-site ordered PSN via synchrotron XRD and TEM imaging. Observation results reveal that the ferroelectric domains have nanoscale meandering shapes, which could be related to the weak relaxor behavior found in this system. Two directions of electric polarization are paired in one set of domains, which is separated from another set of domains by a clear boundary. Although the polarization directions agree with the rhombohedral structure similar to $BiFeO_3$, the wavy pattern and size of the domains in PSN are different from those of other ferroelectric domains reported in the literature. In-situ temperature-variable observation using a high-speed camera directly reveals a ferroelectric phase transition with the coexistence of the ferroelectric and paraelectric phases and domain-wall movements accompanying the release of lattice distortion in the vicinity of the Curie temperature. These results highlight a characteristic ferroelectric nanostructure with both the local polarization fluctuation of a relaxor and the discontinuous phase transition of a normal ferroelectric. Furthermore, the substitution of small amounts of $PbTiO_3$ for PSN changes the wavy ferroelectric domains into microsized domains of straight shapes, while the crystal structure remains in the same $R3m$ space group.

## Acknowledgments

We appreciate the late Dr. Jean-Michel Kiat for his contribution to this work.

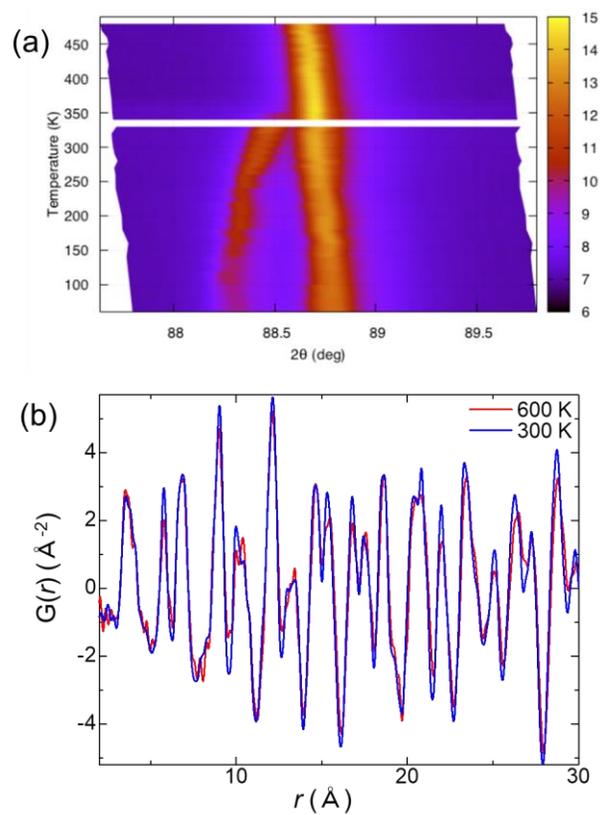

**Figure 1** (a) Temperature dependence of the Bragg reflection $4\bar{4}4$ in ordered PSN. The white line is the blank temperature where the apparatus was changed for heating and cooling. (b) Reduced PDF $G(r)$ at 600 K and 300 K.

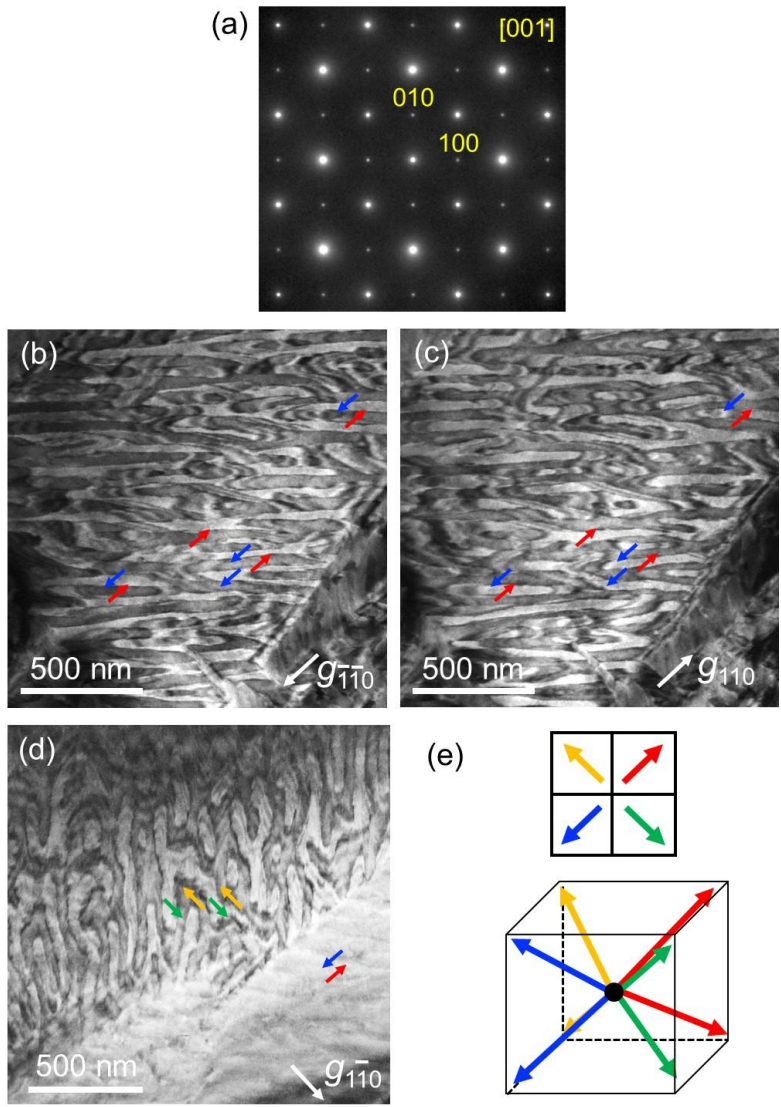

**Figure 2** Observation and analysis of the ferroelectric domains in PSN at room temperature. (a) Electron diffraction pattern along $[001]_{pc}$. (b, c) Dark-field images using $g = \bar{1}\bar{1}0$ and $g = 110$, respectively. (d) Dark-field image of the area next to the region observed in panels (b) and (c). (e) Schematics of the electric-polarization directions for the rhombohedral $R3m$ structure. The arrows correspond to the directions in the dark-field images.

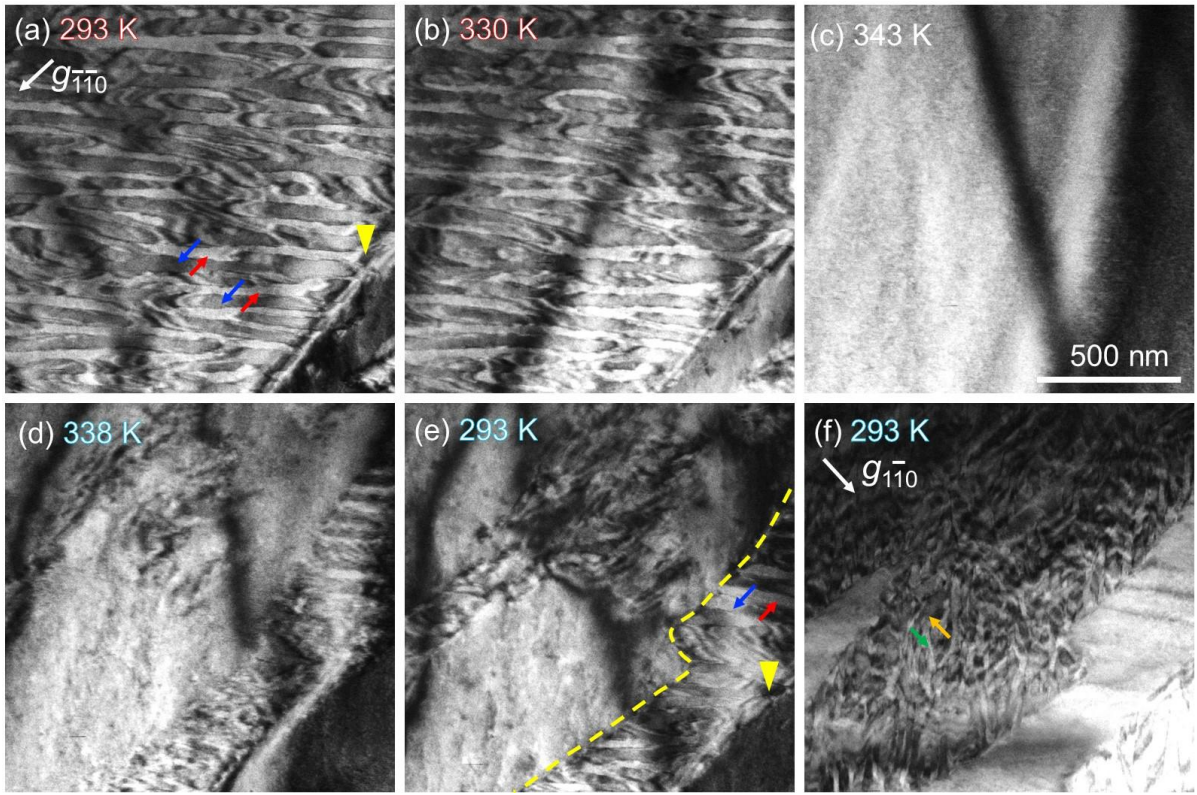

**Figure 3** Dark-field images of the same area in PSN upon heating and cooling. The Bragg reflections $\bar{1}\bar{1}0$ and $1\bar{1}0$ are used for the dark-field imaging in panels (a)–(e) and (f), respectively. The observation plane is $(001)_{pc}$. The yellow line in panel (e) represents the boundary between two sets of domains with different polarization directions. The yellow arrowheads show the position of a crack, where the ferroelectric domains are pinched off. The contrast at 343 K is the mass–thickness contrast and the bend contour.

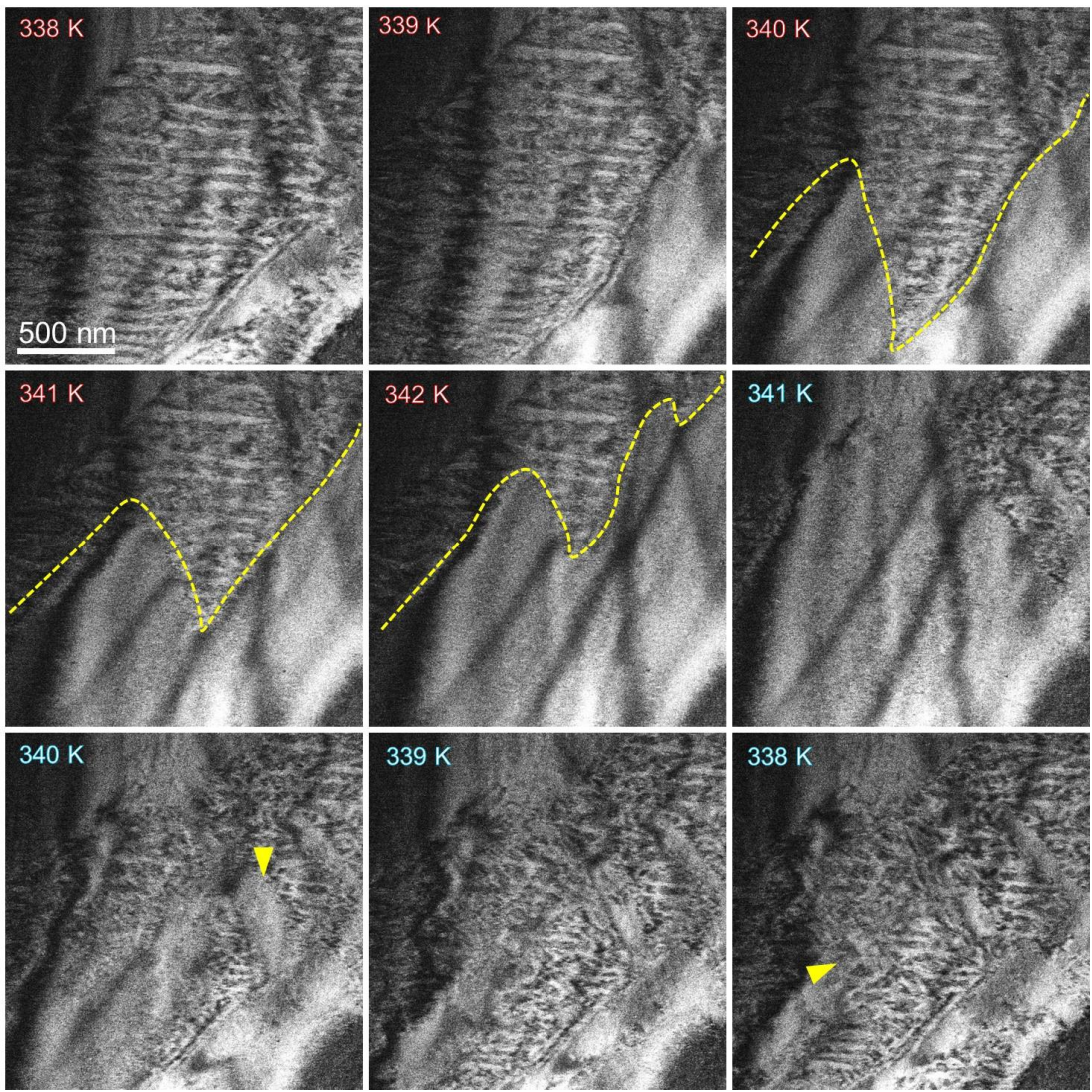

**Figure 4** Observation of the ferroelectric–paraelectric phase transition in the vicinity of the Curie temperature via dark-field imaging while the temperature was changed continuously. The yellow lines represent the boundary between the ferroelectric and paraelectric phases. The arrowheads show the areas where the ferroelectric domains reappeared from the paraelectric regions upon cooling. The transition was captured as a movie, from which these images were reconstructed. See the Supplementary Movie for the entire observation. The Bragg reflection $g = \bar{1}\bar{1}0$ is used for the dark-field imaging, and the observation plane is $(001)_{pc}$.

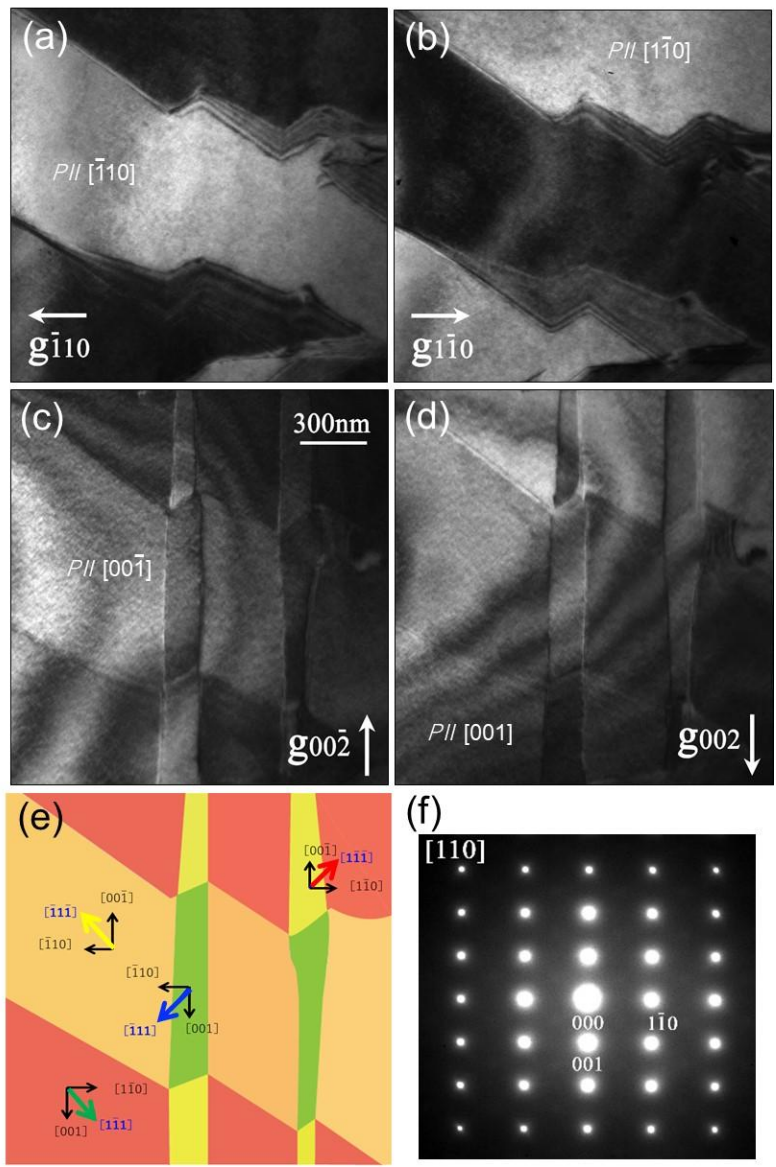

**Figure 5** Dark-field images of 0.80PSN–0.20PT. The Bragg reflections are (a) $g = \bar{1}10$, (b) $g = 1\bar{1}0$, (c) $g = 00\bar{2}$, and (d) $g = 002$. The contrast represents the ferroelectric domains because of the breakdown of Friedel's law. (e) Schematic of the domain orientations. The arrows represent the polarization directions. (f) Electron diffraction pattern along $[110]_{pc}$ from the observed area.

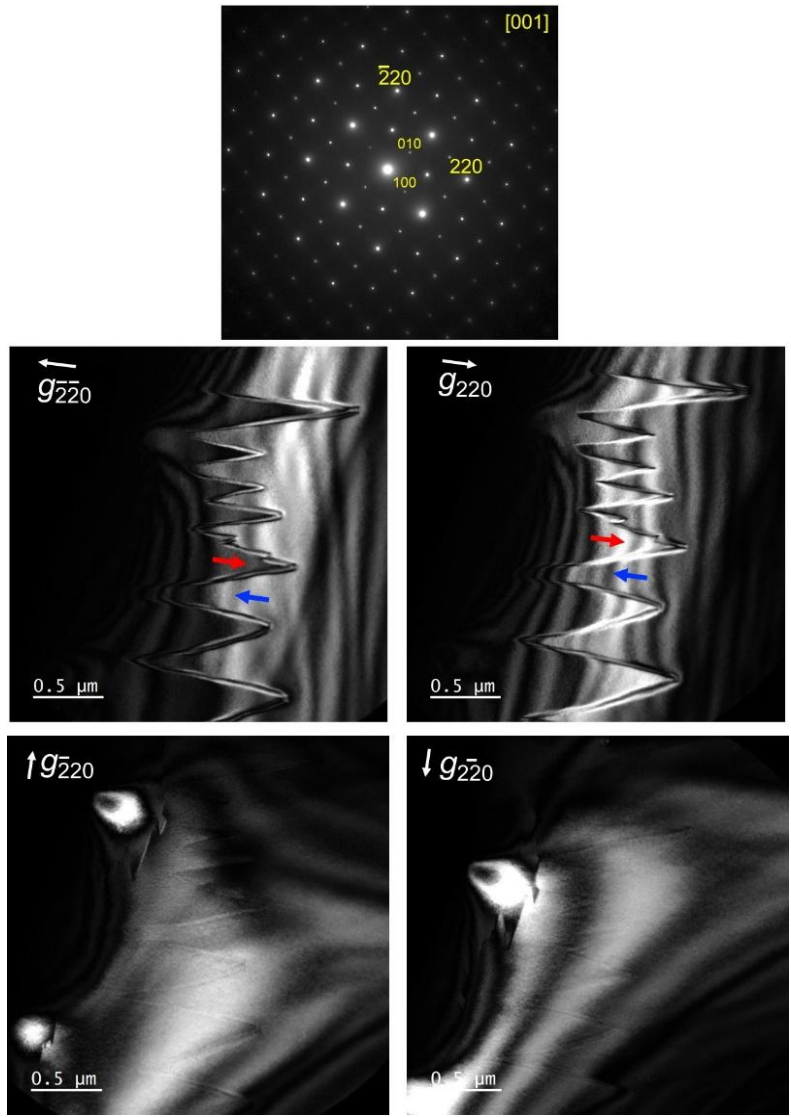

Supplementary Figure S1. Domains observed on a (001) single crystal of $Pb(Sc_{1/2}Nb_{1/2})O_3$ without carbon coating. The mark *g* represents the Bragg reflection used for dark-field imaging.

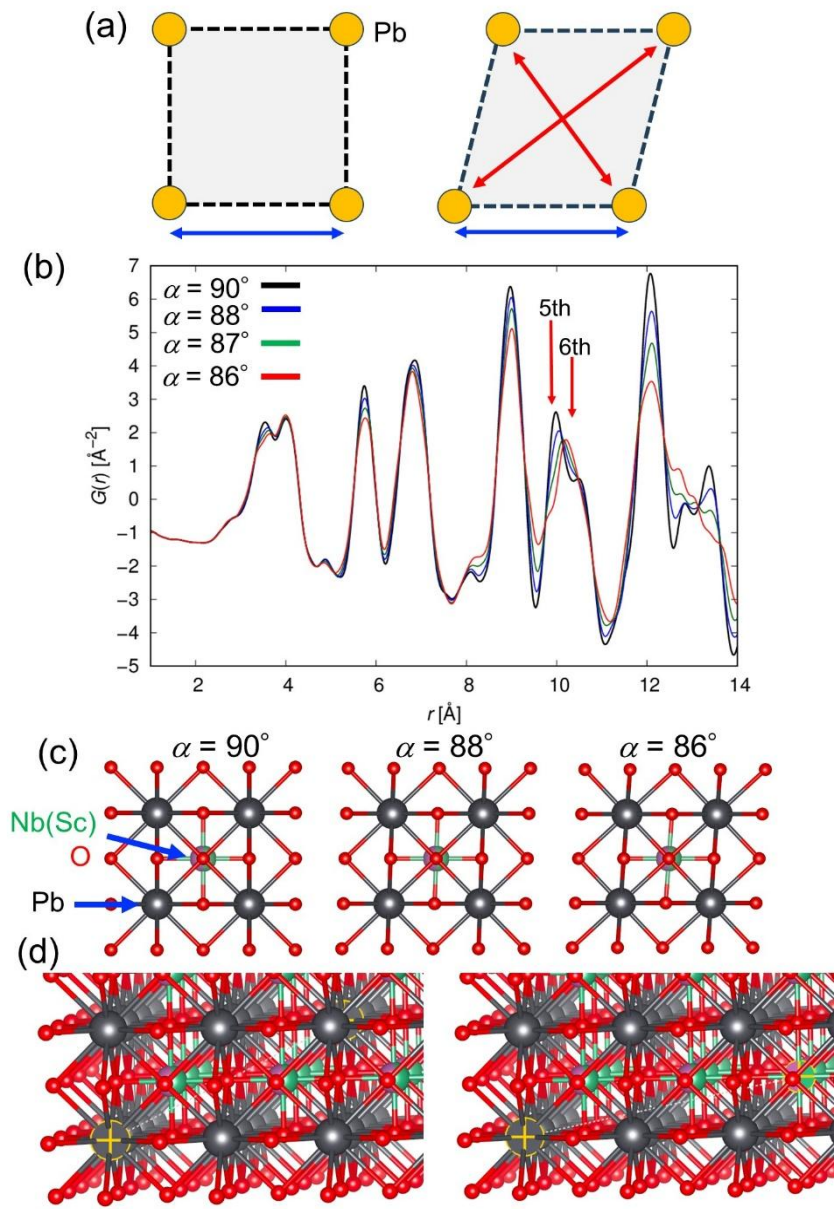

Supplementary Figure S2. (a) Schematic of the primitive unit cell before and after distortion. The blue and red arrows represent the Pb-Pb distances between the nearest neighbors and diagonal directions. (b) Simulation results of pair-distribution function (PDF) when a uniform rhombohedral distortion occurs in the angle $\alpha = 90°$, $88°$, $87°$, and $86°$. (c) Crystal structures with the rhombohedral distortion to calculate the PDF analysis of panel (b). (d) The atomic pairs of (left panel) the fifth (9.98 Å) and (right panel) sixth (10.20 Å) peaks. The fifth and sixth peaks represent Pb–Pb and Pb–Nb(Sc) pairs, respectively, which are marked by the yellow symbols.

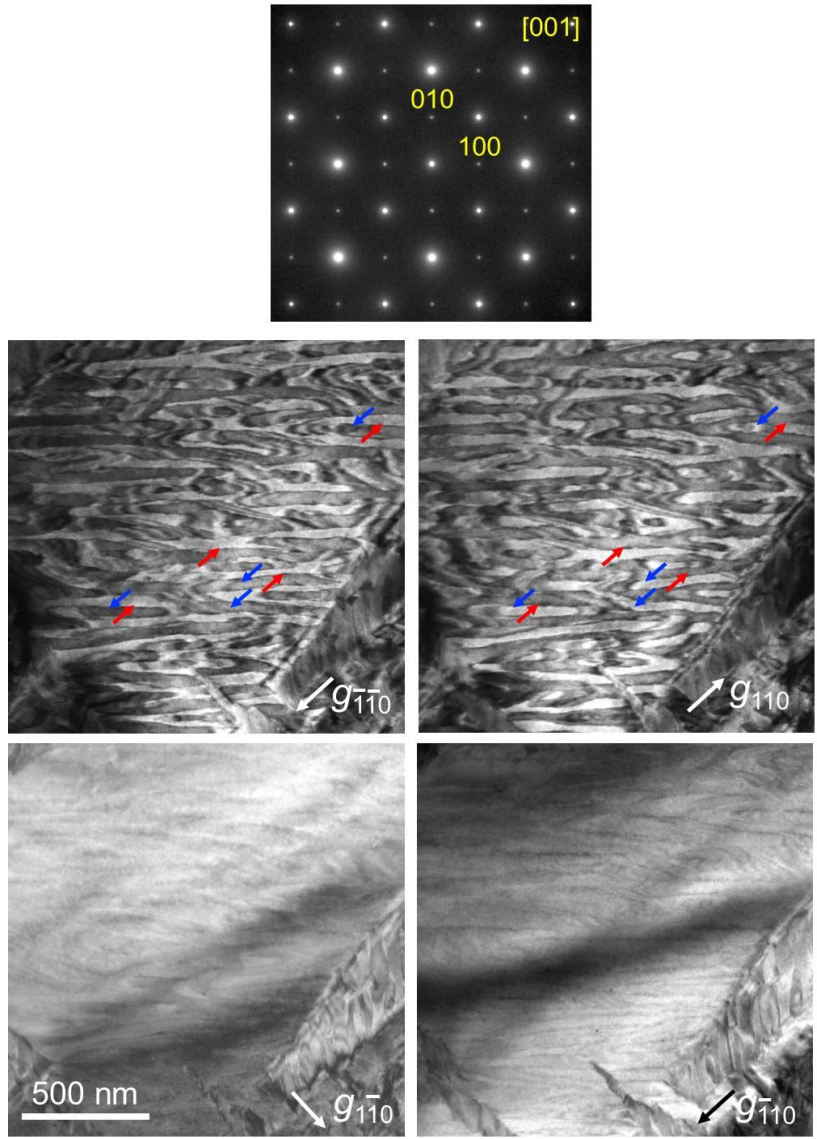

Supplementary Figure S3. Reflection dependence of dark-field images in Pb(Sc$_{1/2}$Nb$_{1/2}$)O$_3$. The four images were captured from the same area.

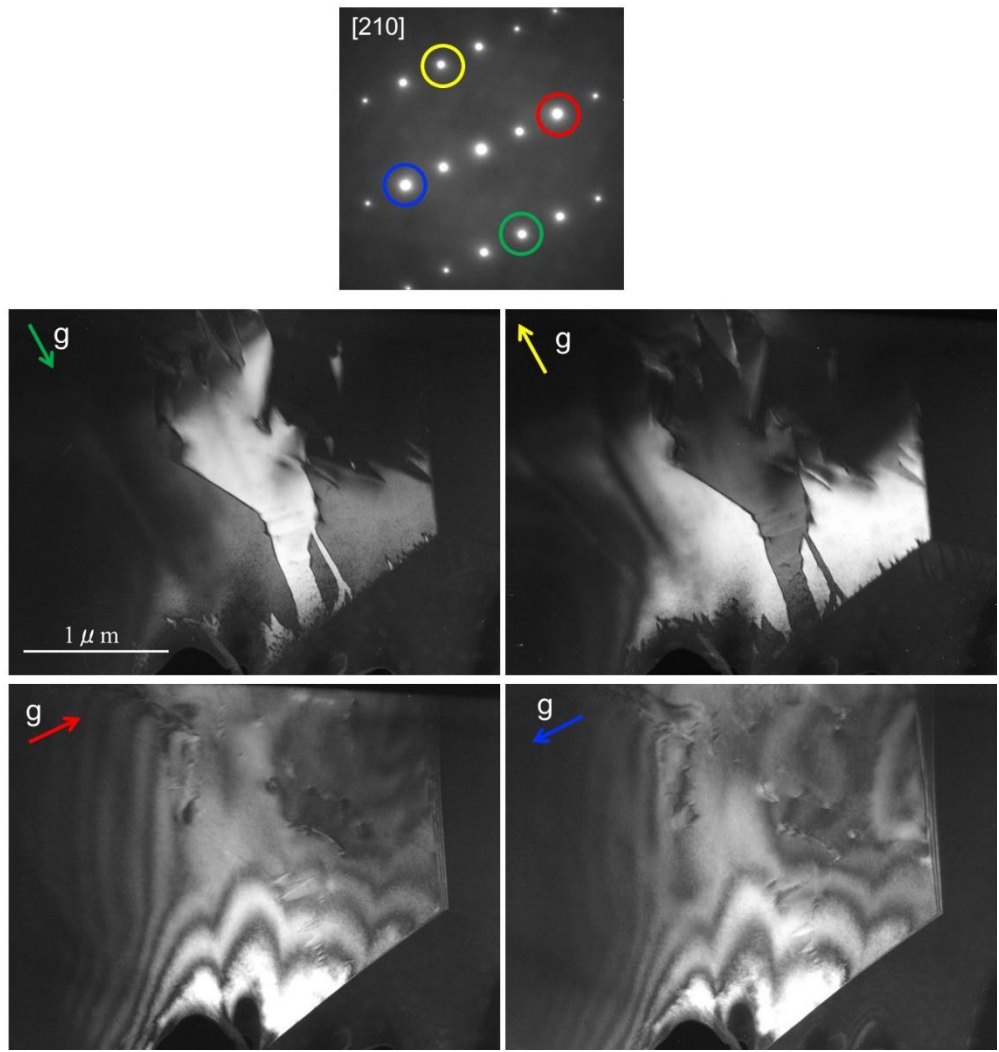

Supplementary Figure S4. Dark-field images in 0.91Pb(Sc$_{1/2}$Nb$_{1/2}$)O$_3$-0.09PbTiO$_3$. The contrast is reversed when the direction of the Bragg reflections is reversed. Besides, the lower panels show the contrast disappearance, demonstrating that the contrast originates from the ferroelectric domains. The colored circles represent the Bragg reflections used for the corresponding dark-field images.

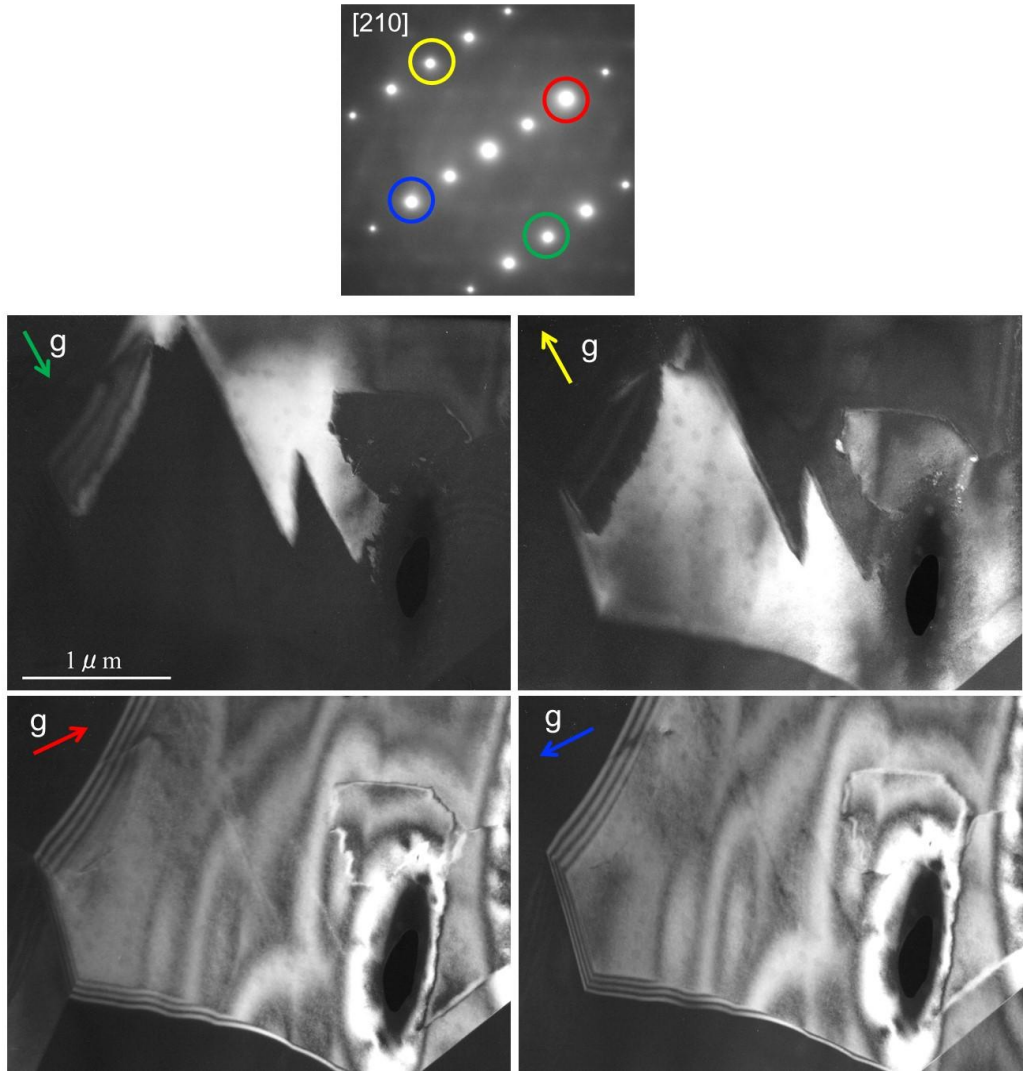

Supplementary Figure S5. Ferroelectric domain observation in 0.91Pb(Sc$_{1/2}$Nb$_{1/2}$)O$_3$-0.09PbTiO$_3$. The observation area is another grain that is different from Supplementary Fig. S4. The colored circles represent the Bragg reflections used for the corresponding dark-field images.